\documentclass[
aps, prl, amsfonts, amssymb, amsmath, floatfix, superscriptaddress, reprint, noeprint, longbibliography
]{revtex4-2} 
\usepackage{graphicx,natbib}
\usepackage{dcolumn}
\usepackage{bm}
\usepackage[breaklinks=true,colorlinks=true,citecolor=blue,urlcolor=blue,linkcolor=blue,pdfencoding=auto, psdextra]{hyperref}
\usepackage{xcolor}
\usepackage{float}
\usepackage{placeins}
\usepackage[T1]{fontenc}
\usepackage[english]{babel}
\usepackage{microtype}

\newcommand{\prlsection}[1]{\emph{#1.}---}
\newcommand\bb[1]{\mbox{\boldmath{$#1$}}}

\begin{document}
\title{Leaking Outside the Box: Kinetic Turbulence with Cosmic-Ray Escape}
\author{Evgeny A.\ Gorbunov}
\email{genegorbs@gmail.com}
\affiliation{Centre for mathematical Plasma Astrophysics, Department of Mathematics, KU Leuven, Celestijnenlaan 200B, B-3001 Leuven, Belgium}
\author{Daniel Gro\v{s}elj}
\affiliation{Centre for mathematical Plasma Astrophysics, Department of Mathematics, KU Leuven, Celestijnenlaan 200B, B-3001 Leuven, Belgium}
\author{Fabio Bacchini}
\affiliation{Centre for mathematical Plasma Astrophysics, Department of Mathematics, KU Leuven, Celestijnenlaan 200B, B-3001 Leuven, Belgium}
\affiliation{Royal Belgian Institute for Space Aeronomy, Solar-Terrestrial Centre of Excellence, Ringlaan 3, 1180 Uccle, Belgium}
\begin{abstract}
We study particle acceleration in strongly turbulent pair plasmas using novel 3D Particle-in-Cell simulations, featuring particle injection from an external heat bath and diffusive escape. We demonstrate the formation
of steady-state, nonthermal particle distributions with maximum energies reaching the Hillas limit. 
The steady state is characterized by the equilibration
of plasma kinetic and magnetic pressures,
which imposes upper limits on the acceleration rate.
With growing cold plasma magnetization $\sigma_0$, nonthermal power-law spectra become harder, and
the fraction of energy channeled into escaping cosmic rays increases.
At $\sigma_0 \gtrsim 1$, the escaping cosmic rays amount to more than 50\% of the dissipated energy.
Our method allows for kinetic studies of particle
acceleration under steady-state conditions, with applications to a variety of astrophysical systems.
\end{abstract}
\maketitle
\prlsection{Introduction}Turbulence is believed to be a widespread mechanism for nonthermal particle acceleration in collisionless plasmas \cite{Fermi_1949,Fermi_1954,kulsrud1_1971,ramaty_1979,dermer_1996,Liu_2004,Brunetti_2007,Petrosian_2008,Sullivan_2009,petrosian_2012,Lynn_2014,Kimura_2016,Sciaccaluga_2022,Zhdankin_2022}.
In recent years, kinetic plasma simulations have provided first-principles insight into the mechanisms of 
particle acceleration in turbulent astrophysical environments by uncovering the nature of 
particle dynamics over their entire acceleration history \cite{Zhdankin_2017_b,zhdankin2017,Zhdankin_2018,zhdankin_2019,Comisso_2019,Wong_2020,Zhdankin_2021,Vega_2022,Vega_2023,Comisso_2024,groselj_2024,Vega_2024}. It was shown that particles can be rapidly
injected from the thermal pool into the suprathermal population by nonideal electric fields
near reconnecting current sheets 
\cite{Comisso_2018_b,Comisso_2019,Sironi_2022,Comisso_2024}. The injection is followed by 
stochastic energization
via interactions with turbulent structures
\cite{Bresci_2022,Pezzi_2022,Kempski_2023,Lemoine_2023,Xu_2023,Vega_2024}. This can be seen as a generalization of the 
second-order Fermi process \cite{Fermi_1949, Fermi_1954,Lemoine_2019,lemoine_2025}, which involves
diffusion in energy space. 

However, most kinetic simulations to date face a significant shortcoming: they lack a genuine steady state, due to the fact that
the simulation domains employed are of finite size and closed, which leads to a progressive accumulation of energy over time and to the
pile-up of particles at the highest energies \cite{Zhdankin_2018,Lemoine2020,Wong2025}. This presents 
challenges for connecting simulations and realistic astrophysical objects, which may convert a significant fraction of their available power
into escaping cosmic rays. Moreover, it is known that 
power-law distributions emerge most naturally from a competition between
particle acceleration and escape. 
In closed domains, the local trapping of 
particles provides an effective ``escape'' mechanism \cite{Lemoine2020},
enabling the development of power-law distributions.
However, without an energy sink to balance the turbulent 
forcing, such power-law distributions are only transient states, followed by 
progressive energy pile-up.
Therefore, steady-state simulations, if achievable, could offer a convenient and sorely needed testbed for theoretical models.

In this Letter, we develop an efficient method for achieving steady state in kinetic plasma-turbulence simulations in open domains, and we characterize the emerging steady-state statistics 
of particle acceleration and escape. As an example, we study large-amplitude 3D Alfv\' enic turbulence in electron--positron plasmas. The steady state is characterized by a near-equilibration of plasma kinetic and magnetic pressures. As a result, bulk turbulent motions remain subrelativistic, which 
limits the rate of stochastic particle acceleration. The particle distributions exhibit nonthermal power laws extending up to the
system size (Hillas) limit, where the rate of diffusive particle escape reaches the theoretical maximum (set by the light-crossing time).
We also show that strongly magnetized sources may convert a large fraction of the available power into escaping cosmic rays.
In our simulations, the nonthermal energy fraction of escaping particles grows with the cold plasma magnetization $\sigma_0$ (defined below) and saturates near 70\% at $\sigma_0\gtrsim 100$.

\prlsection{Kinetic Leaky-Box Model}We perform 3D simulations of kinetic plasma turbulence
with particle injection and escape using the Particle-in-Cell (PIC) code {\tt{Tristan-MP v2}} \cite{hayk_hakobyan_2023_7566725}.
Our model mimics an open turbulent box of side $L=2l_{\rm esc}$, coupled to an external heat 
bath at a fixed temperature $T_0$. 
Particles escape the magnetized turbulent accelerator by diffusing over a distance $l_{\rm esc}$
perpendicular to the mean magnetic field $\bb{B}_0 = B_0\hat{\bb{z}}$, while 
at the same time new thermal particles 
are introduced into the domain from the external heat reservoir. 
To implement escape, 
we introduce the particle variables $\delta x$ and $\delta y$, which
track the displacement of each particle from its initial location 
in $x$ and $y$, respectively 
\footnote{{{ 
Particles inserted at $t>0$ are initialized with 
$\delta x = \delta y = 0$ (i.e., the reference point for the displacement is the 
particle's own initial location). For particles in the box at $t=0$ we make an exception.
Their initial $\delta x$ and $\delta y$ are preset to random numbers between 
$-l_{\rm esc}$ and $l_{\rm esc}$, as if those particles had already 
dispersed from their point of origin. }}}.
A 
particle counts as escaped when $\max(|\delta x|, |\delta y|) > l_{\rm esc}$ (see also \cite{groselj_2024}). In order to 
keep the total number of particles fixed, we insert a new 
thermal particle (sampled from a Maxwellian 
at temperature $T_0$) at the location of every escaping particle. 
The inserted particle has the 
same charge and mass as the escaping particle.

Employing
for simplicity an electron--positron composition ($m_{e^-}=m_{e^+}=m$) of mean number density $n_0$, we explore the effect of different cold plasma 
magnetizations $\sigma_0 = B_0^2 / (4\pi n_0 mc^2)$
on particle acceleration and escape.
Several runs with different $\sigma_0 = \{0.2, 1, 4, 10, 20, 80\}$ are performed.
A periodic box of size $L^3$ is initially filled with thermal particles at temperature $T_0$. 
We set $T_0 = mc^2[\sigma_0(2+\sigma_0)]/[3+3\sigma_0]$, which ensures that
the mean energy of the injected particles $\mathrm{E}_0 \sim mc^2\sigma_0$ is close to the
expected steady-state mean particle energy (see Eq.~\eqref{eq:beta} below).
To sustain turbulence, a Langevin antenna drives an external current \cite{tenbarge_2014} at the
largest (box-scale) wavelengths with frequency $\omega_{\rm ant} \approx 0.9\omega_{\rm A}$, slightly off-resonant to the lowest Alfv\' en frequency of the box $\omega_{\rm A} = (2\pi/L) v_{\rm A}$. The Alfv\'en speed $v_{\rm A}= c \sqrt{\sigma/(\sigma + 1)}$ is defined here based on the initial hot magnetization $\sigma = B_0^2/(4\pi w_0 n_0 mc^2) = \sigma_0/w_0$, where $w_0=\mathcal{K}_3(mc^2/T_0)/ \mathcal{K}_2(mc^2/T_0)$ is the initial specific enthalpy and $\mathcal{K}_n(z)$ are $n$-th order modified Bessel functions of the second kind. The decorrelation rate of the antenna is
$\gamma_{\rm ant} = 0.5\omega_{\rm ant}$. We focus on the strong-turbulence regime, where the magnetic perturbation driven by the antenna is comparable to the background magnetic field: $\delta b \equiv \delta B/B_0 \approx 1$. 
The spatial resolution of all simulations presented in this Letter is $960^3$, and the physical box size
$L = 480 d$, where $d =[(1 + 2\sigma_0) mc^2/(4\pi n_0 e^2)]^{1/2} $ is a reference
hot-plasma skin depth based on the expected steady-state kinetic temperature (discussed below). 
We use 16 particles per cell in all runs \footnote{For a convergence check in 
particle number, see e.g.~\cite{Comisso_2019}.}. The simulations are run for 
at least 20 light-crossing times $t_{\rm cross} = l_{\rm esc}/c$.

\prlsection{Energy Balance in Steady State}
Our simulations presented in the following show that large-amplitude kinetic turbulence with particle escape 
settles into a state where the average
kinetic and magnetic pressures are of the same order. A supporting analytical argument
can be provided by considering the balance between turbulent energization and energy losses from
particle escape:
$(\mathrm{E}_{\rm esc} - \mathrm{E}_0)n_0/t_{\rm esc} \simeq \delta b^2 B_0^2/ (4\pi t_0)$,
where $\mathrm{E}_{\rm esc}$ is the mean energy
of escaping particles, $t_{\rm esc}$ is the escape time, and
$t_0$ is the turbulence cascade time.
Assuming for simplicity that particles from the thermal population stream along
tangled magnetic-field lines, thereby diffusing perpendicularly to the mean magnetic
field, we estimate their escape time 
as $t_{\rm esc}\simeq N\delta t\simeq l_{\rm esc}^2/(l_{\parallel}\delta b^2 v_{\rm th})$, 
where $\delta t\simeq l_{\parallel}/v_{\rm th}$ is the time to travel one field-parallel
correlation length $l_{\parallel}$ at
thermal speed $v_{\rm th}\equiv[T/(w m)]^{1/2}$ and $N \simeq [l_{\rm esc} / (l_{\parallel}\delta b)]^2$
is the number of correlation lengths needed to diffuse over a distance $l_{\rm esc}$.
The cascade time $t_0 \simeq \chi^{-2}l_{\parallel}/v_{\rm A}\simeq l_0^2 / (l_{\parallel}\delta b^2 v_{\rm A})$,
where $\chi = \delta b\, l_{\parallel}/l_0$ is the turbulence nonlinearity parameter and
$l_0$ is the perpendicular coherence scale.
Focusing on the regime of interest with $\sigma_0\gtrsim 1$, 
{ $(\mathrm{E}_{\rm esc} - \mathrm{E}_0)/\mathrm{E}_0\gtrsim 1$}, and assuming at most mildly relativistic Alfv\'en speeds, such that
$v_{\rm A}^2 \simeq c^2\sigma_0 / w$ (to be justified below), we finally obtain
\begin{align}
    \beta \sim  (\delta b\,l_{\rm esc}/l_0)^{4/3},
    \label{eq:beta}
\end{align}
where $\beta \simeq 2 v_{\rm th}^2 / v_{\rm A}^2 \simeq 2 T / (\sigma_0 mc^2)$ is the plasma beta, i.e., the ratio of 
plasma kinetic to magnetic pressures. Thus, large-amplitude ($\delta b \sim 1$)
magnetized turbulence, 
driven on scales comparable to
the particle escape distance ($l_0 \sim l_{\rm esc}$), naturally
settles into a state with $\beta\sim 1$. This enforces an upper limit on the hot
plasma magnetization $\sigma\lesssim 1$, thereby limiting the
turbulent motions $\delta v\simeq\delta b v_{\rm A}$ to the
mildly relativistic regime. In turn, this limits the maximum particle scattering rate and the resulting acceleration efficiency,
as shown below.

\prlsection{Simulation Results}In Fig.~\ref{fig:figure1} we demonstrate the formation of a self-consistent steady state in our 
kinetic leaky-box model, using the $\sigma_0 = 20$ case as an example.
The temporal evolution of the 1D magnetic spectrum ${\rm E}_{\rm mag} (k_\perp)$ as
a function of the field-perpendicular wave number $k_{\perp}$
is shown in Fig.~\hyperref[fig:figure1]{\ref*{fig:figure1}(a)}. 
A steady-state turbulence spectrum is obtained after $\approx 5l_{\rm esc}/c$. At magnetohydrodynamic
(MHD) scales ($k_\perp d \ll 1$), the spectrum follows a familiar scaling $\propto k_\perp^{-5/3}$ \cite{goldreich1995}, contrary
to recent theoretical expectations \cite{lemoine2024} arguing that nonlinear cosmic-ray feedback leads to a significant
steepening of the MHD turbulence spectrum.
The formation of steady-state nonthermal particle distributions is depicted in Fig.~\hyperref[fig:figure1]{\ref*{fig:figure1}(b)}.
The initially Maxwellian energy distribution 
develops nonthermal features at high energies, characterized by a power-law dependence $f({\rm E}) \equiv {\rm dN}/{\rm dE} \propto \gamma^{-p}$. For $\sigma_0 = 20$, the power-law index reaches a mean value of $p\approx2.8$ after $\approx7$ light-crossing times. The presence of escape allows us to measure the distribution of the escaping particles $f_{\rm esc}({\rm E})$, which develops a slightly harder power-law tail, saturating at an index $p\approx 2.5$.

\begin{figure}
\centering
\includegraphics[width=1\columnwidth, trim={0mm 0mm 0mm 0mm},clip]{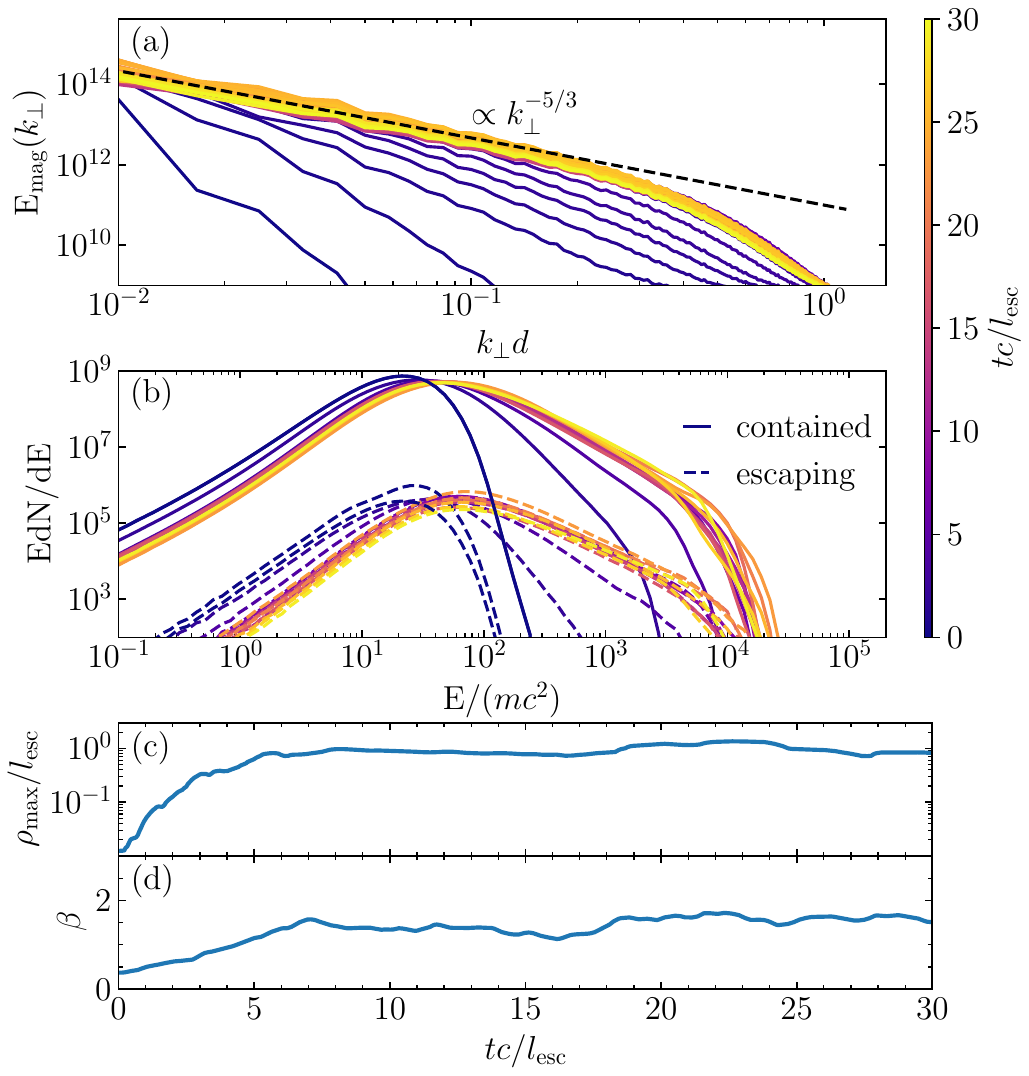}
\caption{Approach toward steady state at $\sigma_0 = 20$.
(a) Evolution of magnetic-field spectra (colors represent time). The dashed line shows a $-5/3$ slope for reference. (b) Temporal evolution of the particle energy spectra. Escaping-particle spectra are shown with dashed lines, contained with solid lines. (c) The Larmor radius
$\rho_{\rm max}(t)$ of the highest-energy particles. (d) Evolution of the
plasma $\beta \equiv 8\pi n_0 T/B_{\rm rms}^2$ over time.
}
\label{fig:figure1}
\end{figure}

In Fig.~\hyperref[fig:figure1]{\ref*{fig:figure1}(c)}, we track the evolution of the Larmor
radius $\rho_{\rm max} = (\gamma_{\rm max}^2 - 1)^{1/2} mc^2/(qB_0)$ of the highest-energy particles 
in the box \footnote{we determine $\gamma_{\rm max} = {\rm E}_{\rm max}/(mc^2) + 1$
based on the highest energy at which ${\rm E}_{\max} f({\rm E}_{\max})$ 
is less than a factor of $10^5$ below the peak of ${\rm E} f({\rm E})$.}. 
As appropriate for $\delta b\sim 1$, we assume that the
highest-energy particles have large pitch angles \citep{Comisso_2019}; for $\delta b\ll 1$ the (small) pitch angle 
should be included in the estimate of $\rho_{\rm max}$ \citep{Nattila_2022,Vega_2025_arx}.
At early times, $\rho_{\rm max}(t)$ {in Fig.~\hyperref[fig:figure1]{\ref*{fig:figure1}(c)}} grows at 
near-exponential rate \citep{Vega_2024}, which is followed by saturation at the 
Hillas limit $\rho_{\rm max}\approx \rho_{\rm Hillas} \approx l_{\rm esc}$.
The energization of the plasma and its approach toward steady state is evident also from the
evolution of the plasma-beta parameter $\beta \equiv 8\pi n_0T/B_{\rm rms}^2$ (Fig.~\hyperref[fig:figure1]{\ref*{fig:figure1}(d)}), 
where $T = [mc^2/(3 n_0)] \int (\gamma - 1/\gamma)f(\rm E){\rm dE}$ and $\gamma \equiv {\rm E}/(mc^2) + 1$ \footnote{For simplicity, and considering the absence of strong bulk flows, we compute the kinetic
and magnetic pressures in the simulation frame, rather than in the proper
fluid frame \cite{Zhdankin_2021}. We confirmed that a more elaborate estimate based on the transformation into the fluid frame leaves results practically unchanged.}.  
From panel (d) of Fig.~\ref{fig:figure1}, we can conclude that the average $\beta\approx1.5$ in steady state. 
This is a generic feature of all of our runs: for any value of the cold magnetization $\sigma_0$, the system eventually relaxes toward a state in which the 
plasma kinetic and magnetic pressures equilibrate, in line with the theoretical argument presented above.
Such a universal behavior at high magnetization introduces an upper limit on the bulk 
turbulent motions of the plasma. Since the hot magnetization $\sigma = \sigma_0/w \approx 0.33$ (where we used $w \approx 4 T/ (mc^2)\approx 3\sigma_0$), the typical Alfv\'en speed is $v_{\rm A}\approx 0.5c$, implying that the bulk motions are at most mildly relativistic, at least in the absence of radiative-cooling mechanisms \cite{zhdankin2017, Vega_2022, groselj_2024}.

\begin{figure}[t]
\centering
\includegraphics[width=1\columnwidth, trim={0mm 0mm 0mm 0mm},clip]{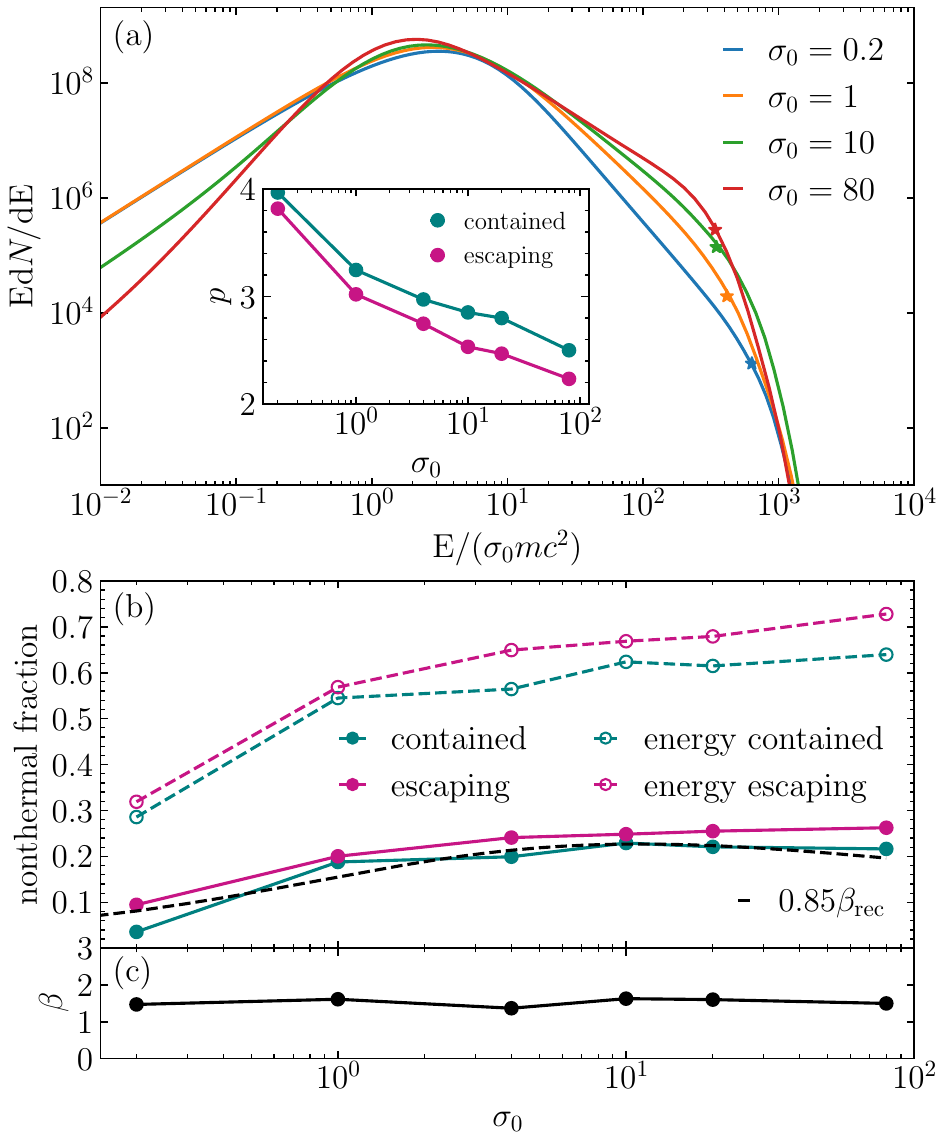}
\caption{(a) Steady-state particle distributions of the contained population. Stars 
denote the Hillas limit. The inset shows the measured power-law indices for both contained (green line) and escaping (pink line) particle populations. (b) The number fraction of nonthermal particles (solid lines) and 
the fraction of energy attributed to nonthermal particles (dashed lines) for the 
contained (green) and escaping (pink) population.
(c) The measured steady-state plasma $\beta$ for different cold magnetizations $\sigma_0$. 
{For simulations with $\delta b < 1$ see Appendix A.}}
\label{fig:figure2}
\end{figure}

How the plasma magnetization $\sigma_0$ shapes the nonthermal particle distributions
is shown in Fig.~\ref{fig:figure2}. Fig.~\hyperref[fig:figure2]{\ref*{fig:figure2}(a)} shows the steady-state particle distributions as a function of ${\rm E}/(\sigma_0mc^2)$. Due to the linear dependence $T\propto \sigma_0mc^2$ (see Eq.~\eqref{eq:beta}), 
the distribution peaks 
approximately overlap, albeit with a slight shift toward lower energies for higher $\sigma_0$. Efficient acceleration is enabled
in all cases considered, in the sense that all distributions form extended nonthermal tails reaching the
maximum energy set by the system-size (Hillas) limit ($\rho_{\rm max}\approx l_{\rm esc}$; marked with star symbols
in panel (a)). More strongly magnetized systems feature harder nonthermal tails, as shown in the inset of 
Fig.~\hyperref[fig:figure2]{\ref*{fig:figure2}(a)}. The obtained power-law slopes are broadly 
consistent with previous kinetic simulations of strong turbulence in 
closed domains \cite{Zhdankin_2017_b,Comisso_2018_b,Comisso_2019,Vega_2022,Vega_2023}.

In Fig.~\hyperref[fig:figure2]{\ref*{fig:figure2}(b)} we show the fraction of particles and of the kinetic energy
contained in nonthermal (i.e., non-Maxwellian) particles.
The nonthermal population is defined as $f_{\rm nt} = f({\rm E}) - f_{\rm M}(\rm E)$, where $f_{\rm M}(\rm E)$ is 
a Maxwellian distribution fitted with the data below the peak of $f(\rm E)$. The fraction of 
energy contained in nonthermal particles exceeds 50\% at $\sigma_0\gtrsim 1$. Therefore, our results show that
strongly magnetized turbulent accelerators can release a large fraction of the dissipated power into escaping
cosmic rays (i.e., into nonthermal particles). 
At $\sigma_0\gtrsim 100$ the escaping cosmic rays amount to about $\approx 20$\% of the
particles leaving the box and they carry away $\approx 70$\% of the dissipated turbulence power.
Finally, we confirm that the steady-state plasma $\beta$ is remarkably independent of 
$\sigma_0$ and maintains a value $\beta\approx 1.5$ in all simulations, 
as shown in Fig.~\hyperref[fig:figure2]{\ref*{fig:figure2}(c)}, in line with our analytical expectation.

In strongly magnetized turbulence, particles can be injected into the nonthermal population 
at reconnecting current sheets \cite{Comisso_2019}. Following the reconnection scenario, we can estimate the fraction of nonthermal 
particles $\xi_{\rm nt }$ from the expected
rate of reconnection in relativistic plasmas \citep{Goodbred_2022}.
In one turbulence eddy turnover time, reconnecting sheets process a fraction 
${\mathcal V}_{\rm rec}\sim \beta_{\rm rec}$ of the plasma volume \citep{Comisso_2019}, where $\beta_{\rm rec}$ is the reconnection
rate in units of $B_0v_{\rm A}$, assuming $\delta B \sim B_0$ 
(i.e., moderate-guide-field reconnection). The reconnection rate can be obtained from the slope $S$
of the separatrix field lines (defining the reconnection exhaust) as 
$\beta_{\rm rec} \simeq S(1-S^2)/(1+S^2)$ \cite{liu_2017,Liu2022,Goodbred_2022}. {
Following \citep{Goodbred_2022}, we express \footnote{{
Eq.~\eqref{eq:rate} is formally derived assuming a cold plasma 
upstream of the reconnection layer \citep{Goodbred_2022}. We adopt it as a crude estimate, 
given that a more precise expression is not known to us.}}:}
\begin{equation}
    S^2 = 1 - \frac{[2+\sigma_{\rm m}/2](1+S^2)/(1-S^2) - 1}{[1+\sigma_{\rm m}/2]\{1+(\sigma_{\rm m}/2)(1-S^2)/(1+S^2)\}^{1/2}},
    \label{eq:rate}
\end{equation}
where $\sigma_{\rm m}\approx \sigma_0 (1 - S^2)/(1 + S^2)$ is the ``microscale'' magnetization
parameter, and we took half of the asymptotic X-line kinetic pressure 
to obtain the above formula \citep{Goodbred_2022}. Assuming that an order-unity fraction of particles reprocessed by reconnection 
is injected into the acceleration process \citep{Hoshino_2023}, we finally estimate the 
nonthermal fraction as $\xi_{\rm nt}\sim \beta_{\rm rec}$. 
Our estimate $\xi_{\rm nt}\approx 0.85 \beta_{\rm rec}$ (with 0.85 as an ad hoc prefactor) is shown with a black dashed line 
in Fig.~\hyperref[fig:figure2]{\ref*{fig:figure2}(b)}, demonstrating good agreement with the simulations. 
{ We note that particles need not be injected exclusively 
by the nonideal electric fields at reconnecting current sheets \citep{Singh_2025}; different forms of injection 
are still debated (e.g., \citep{Sironi_2022,Guo2023}). In particular, with decreasing $\sigma_0$
the effect of nonideal fields is expected to diminish, as the particles become less magnetized.}

\begin{figure}[t]
\centering
\includegraphics[width=1\columnwidth, trim={0mm 0mm 0mm 0mm}, clip]{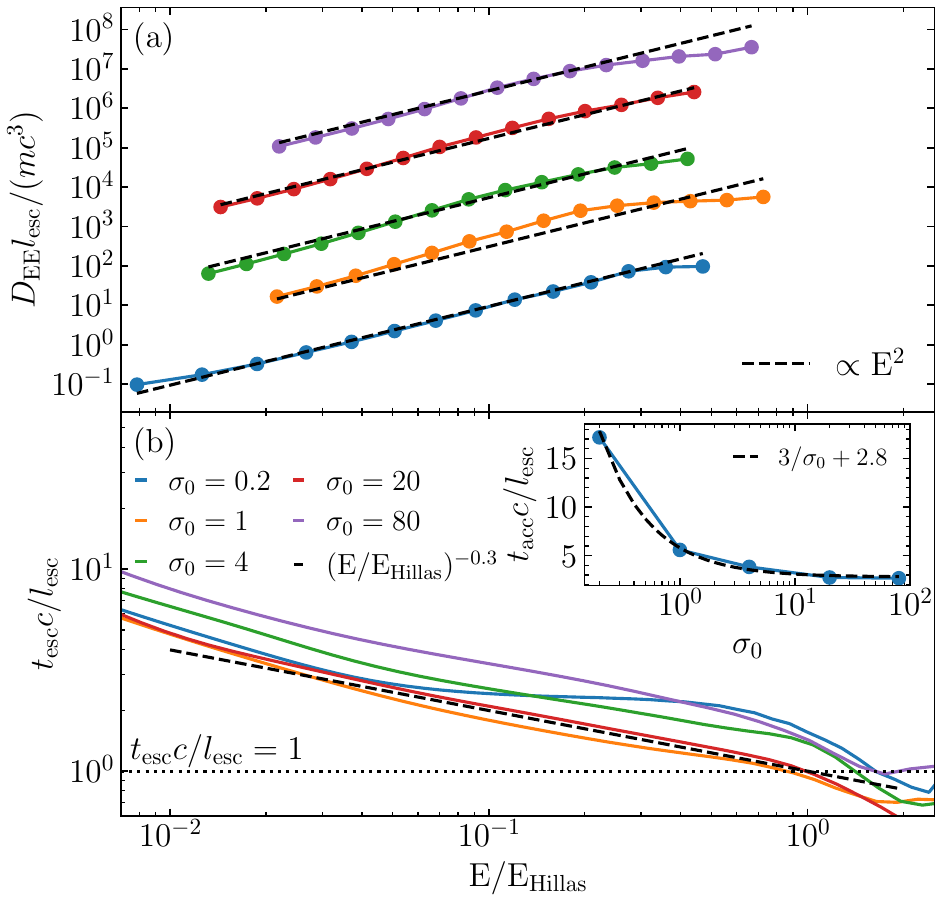}
\caption{(a) Energy diffusion coefficient versus energy for different $\sigma_0$. Dashed black lines indicate 
fits of the form $D_{\rm EE} = {\rm E}^2/t_{\rm acc}$, where $t_{\rm acc}$ is the
acceleration time. (b) Escape time $t_{\rm esc}$ versus energy for different $\sigma_0$. 
The inset shows the acceleration time $t_{\rm acc}(\sigma_0)$, which is well-described by a
fit of the form $t_{\rm acc}c/l_{\rm esc} \approx 3/\sigma_0 + 2.8$.
}
\label{fig:figure3}
\end{figure}

In Fig.~\ref{fig:figure3}, we study the statistics of particle acceleration and escape. We obtain the acceleration 
time scale $t_{\rm acc}$ by measuring the energy 
diffusion coefficient $D_{\rm EE}(\rm E)$ \citep{Comisso_2019,Wong_2020}, which defines the acceleration
time as $t_{\rm acc}(\rm E) = \rm {E}^2/D_{\rm EE}(\rm E)$.
To this end, we track simulation particles located in different energy bins
starting from a time $t_*$ 
\footnote{The particle energies are computed in the frame comoving with the drift velocity $c \bb{E}\times \bb{B}/B^2$ 
and averaged over the gyration time, to remove high-frequency oscillations.}, and 
we obtain the energy-diffusion coefficient
as $D_{\rm EE}({\rm E}) = \langle\Delta {\rm E}(t)^2\rangle/\Delta t$,
where $\Delta {\rm E}(t) = {\rm E}(t) - {\rm E}(t_*)$, $\Delta t = t - t_*\approx l_{\rm esc}/c$, and
the angular brackets represent an average over all particles in a given energy bin.
The dependence of diffusion coefficients on energy for different magnetizations is shown in Fig.~\hyperref[fig:figure3]{\ref*{fig:figure3}(a)}. The diffusion coefficients scale as $D_{\rm EE}\propto {\rm E}^2$, consistent with 
previous measurements performed in closed domains \cite{Comisso_2019,Wong_2020}. A slight departure from the 
$\propto {\rm E}^2$ scaling is seen at high energies, approaching the Hillas limit. 
By performing fits of the form $D_{\rm EE}(\rm E) = E^2/t_{\rm acc} $ (dashed black lines), we determine 
the acceleration time $t_{\rm acc}(\sigma_0)$ (inset of Fig.~\hyperref[fig:figure3]{\ref*{fig:figure3}(b)}). 
The efficiency of acceleration is related to the speed of the scattering agents \cite{Lemoine_2019}, which typically move with the velocity of the turbulent bulk flow $\delta v \sim v_{\rm A}$. The rate of stochastic acceleration is then limited by the largest achievable steady-state bulk velocity. This is reflected in our fitting formula $t_{\rm acc} \approx (3/\sigma_0 + 2.8)l_{\rm esc}/c$. 
For $\sigma_0\gg1$, the acceleration time settles at $t_{\rm acc} \approx 2.8 l_{\rm esc}/c$, which represents the 
asymptotic value set by the heating-regulated ceiling for the Alfv\'en speed.

We define the escape time based on the energy-dependent flux of escaping particles (e.g., \cite{Becker_2006}) as
$t_{\rm esc} = f({\rm E})/\dot{f}_{\rm esc}({\rm E})$, where $\dot{f}_{\rm esc}({\rm E})$ is the number of particles
leaving the box per unit of energy and per unit time.
The measured $t_{\rm esc}({\rm E})$ are shown in Fig~\hyperref[fig:figure3]{\ref*{fig:figure3}(b)}. An approximate
fitting formula is $t_{\rm esc}\approx {({\rm E}/{\rm E}_{\rm Hillas})^{-0.3}}l_{\rm esc}/c$. The $-0.3$ exponent is 
consistent with existing literature \cite{Kempski_2023, Lemoine_2023, Comisso_2024}. Note that for the moderate-magnetization case ($\sigma_0=0.2$), 
$t_{\rm esc}$  is instead approximately constant in the nonthermal tail. A lower limit on the
escape time is imposed by the light-crossing time $l_{\rm esc}/c$
(dotted horizontal line in Fig.~\hyperref[fig:figure3]{\ref*{fig:figure3}(b)}). We observe that this 
limit is approximately reached at the highest energies, corresponding to the Hillas limit, 
which reflects the fact that the highest-energy particles are not confined by the
accelerator and thus escape via ballistic motion. 

\prlsection{Discussion}We conducted the first PIC simulations of kinetic turbulence with diffusive particle escape. Our method
allows us to study the self-consistent production of cosmic rays in \textit{true} steady-state turbulence, which was never achieved before in fully kinetic simulations of turbulent plasmas. Our novel approach has potential applications to a variety of
moderately to strongly magnetized astrophysical objects, such as
pulsar wind nebulae \cite{Porth2014,Lyutikov2019,Xu2019,Comisso2020},
jets from supermassive black holes \cite{MacDonald2018,Alves_2018,Davelaar_2020}, and accretion disks in active galactic nuclei \cite{Bacchini_2022,bacchini_2024,Gorbunov_2025}.
As an example, we focused on 3D simulations employing a pair-plasma composition.
We have demonstrated the formation of steady-state nonthermal particle distributions featuring extended power-law tails reaching the
system size (Hillas) limit. The nonthermal power-law index $p$ hardens with the cold plasma magnetization $\sigma_0$.
At $\sigma_0\gg 1$, our results indicate an asymptotic value of $p\approx 2.3$ and $p\approx 2$ for the contained 
and escaping particle populations, respectively.
The steady state is characterized by the equilibration of plasma kinetic and
magnetic pressures ($\beta\approx 1$).
This effect constrains the magnitude of turbulent bulk motions to subrelativistic 
speeds, which imposes limits on the maximum acceleration rate.
This means that the acceleration time under steady-state conditions is bounded from below when $\sigma_0\gg 1$. 
In our simulations, the lower bound is at $t_{\rm acc} \approx 2.8 l_{\rm esc}/c$. We also measure the
energy-dependent escape time and find that $t_{\rm esc}\approx {({\rm E}/{\rm E}_{\rm Hillas})^{-0.3}}l_{\rm esc}/c$.

Our numerical experiments also show that strongly magnetized turbulent accelerators can release a large fraction of the
dissipated power into escaping cosmic rays, which can then feedback on the ambient medium.
At $\sigma_0\gg 1$, the nonthermal particles (i.e., cosmic rays) carry up to $\approx 70$\% of the total escaping energy flux. Such a
large amount of energy contained in the nonthermal tail might provide nonlinear feedback on the turbulent cascade itself \cite{lemoine2024}; contrary to expectations, we do not observe any significant modification of the turbulence 
spectrum in our simulations, and this aspect warrants further investigation.

\begin{acknowledgments}
E.A.G.\ acknowledges useful discussions with M. Lemoine, A. Philippov, and V. Zhdankin during the development of this work. D.G.\ thanks L.~Comisso and
L.~Sironi for stimulating discussions, which motivated this study.
F.B.\ acknowledges support from the FED-tWIN programme (profile Prf-2020-004, project ``ENERGY'') issued by BELSPO, and from the FWO Junior Research Project G020224N granted by the Research Foundation -- Flanders (FWO).
D.G.~is supported by the Research Foundation -- Flanders (FWO) Senior Postdoctoral  Fellowship 12B1424N.
The resources and services used in this work were provided by the VSC (Flemish Supercomputer Center), funded by the Research Foundation - Flanders (FWO) and the Flemish Government.
\end{acknowledgments}

\FloatBarrier
\bibliography{literature}
\newpage
\appendix
\section*{End Matter}
\begin{figure}[h]
\centering
\includegraphics[width=1\columnwidth, trim={0mm 0mm 0mm 0mm},clip]{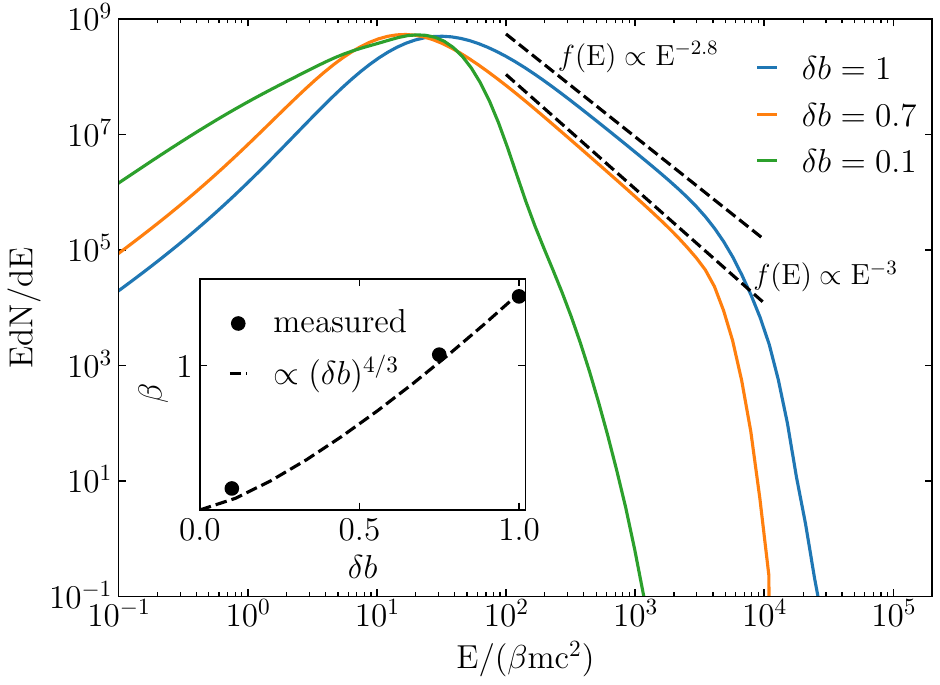}
\caption{ Steady-state particle distributions at different $\delta b$. The inset shows the dependence of plasma $\beta$ on the strength of magnetic-field fluctuations $\delta b$. The dashed black curve demonstrates the fit by our analytical formula \eqref{eq:beta}.}

\label{fig:db}
\end{figure}

 \prlsection{Appendix A: Dependence on $\delta b$}We performed additional simulations at $\sigma_0 = 20$ to check the 
dependence on the 
turbulence magnetic fluctuation strength $\delta b$. We scale $T_0$ in proportion to $\delta b^2$, so that $T_0$ 
is always below the expected steady-state temperature of the turbulent plasma.
As seen from Fig.~\ref{fig:db}, the efficiency of the nonthermal acceleration is significantly reduced at $\delta b<1$, 
in agreement with previous works \cite{Comisso_2018_b,Vega_2022,
Nattila_2022, Vega_2023,Vega_2024}. We also compute the steady-state plasma $\beta$ for these runs and compare it to 
our prediction~\eqref{eq:beta}. The result is shown in the inset of Fig.~\ref{fig:db}. We note the good agreement between our analytic prediction and the simulations.   
\end{document}